# Observation of Broken Time-Reversal Symmetry in the Heavy Fermion Superconductor UPt$_3$

E. R. Schemm[1,2,3]*, W. J. Gannon[4], C. M. Wishne[4], W. P. Halperin[4], A. Kapitulnik[1,2,3,5]

**Affiliations:**

[1]Department of Physics, Stanford University, Stanford, CA 94305, USA.

[2]Stanford Institute for Materials and Energy Sciences (SIMES), SLAC National Accelerator Laboratory, 2575 Sand Hill Road, Menlo Park, CA 94025, USA

[3]Geballe Laboratory for Advanced Materials, Stanford University, Stanford, CA 94305, USA.

[4]Department of Physics and Astronomy, Northwestern University, Evanston, IL 60208, USA.

[5]Department of Applied Physics, Stanford University, Stanford, CA 94305, USA.

*Correspondence to: eschemm@stanford.edu

**Abstract**: The symmetry properties of the order parameter characterize different phases of unconventional superconductors. In the case of the heavy-fermion superconductor UPt$_3$, a key question is whether its multiple superconducting phases preserve or break time-reversal symmetry (TRS). We tested for asymmetry in the phase shift between left and right circularly polarized light reflected from a single crystal of UPt$_3$ at normal incidence, finding that this so-called polar Kerr effect appears only below the lower of the two zero-field superconducting transition temperatures. Our results provide evidence for broken TRS in the low-temperature superconducting phase of UPt$_3$, implying a complex two-component order parameter for superconductivity in this system.

***************************

The heavy-fermion metal UPt$_3$ (*1*) is one of only a handful of unconventional superconductors (*2*, *3*) exhibiting multiple superconducting phases (*4-6*). In the normal state, strong hybridization between itinerant platinum 5$d$ electrons and localized uranium 5$f$ moments results in an effective mass that is ~50 times that of free electrons (*2*). Below the Néel temperature $T_N \sim 5$ K, the local U moments order antiferromagnetically in the *a-b* plane (*7*). In zero magnetic field, two peaks in the specific heat at $T_{c+} \sim 550$ mK and $T_{c-} \sim 480$ mK indicate the presence two superconducting states of differing symmetry, called the A and B phases respectively (*4-6*). Pressure studies suggest that these two phases couple to and are stabilized by the antiferromagnetic order parameter (*8*). In finite magnetic fields, three distinct vortex phases are also observed (*9-11*). The phase diagram of UPt$_3$ therefore presents a unique challenge for models of unconventional superconductivity.

In the absence of a detailed understanding of the microscopic origins of unconventional superconductivity, theoretical and experimental efforts center on identifying the structure of the macroscopic superconducting order parameter, i.e. the pair wavefunction or gap. In the case of UPt$_3$, acceptable candidate order parameters should respect the $D_{6h}$ point-group symmetry of the underlying crystal lattice and should therefore transform under one or more representations of this group. In this framework, many – but not all – experimental studies of the superconducting states (*1*) favor an $E_{2u}$ odd-parity triplet representation in which the gap is given by

$$\Delta(\mathbf{k}_F) = \hat{z}\left[\eta_1(T)\left(k_x^2 - k_y^2\right)k_z \pm 2i \cdot \eta_2(T)k_x k_y k_z\right] \quad (1)$$

in a coordinate system where $\hat{z} \parallel c$. Here $\eta_1(T) \propto \sqrt{1-(T/T_{c+})^2}$ is the real component of the superconducting order parameter, marking the onset of the A phase at $T_{c+}$, while $\eta_2(T) \propto \sqrt{1-(T/T_{c-})^2}$ introduces an additional imaginary component in the B phase at $T_{c-}$ (*12, 13*). An order parameter of this form first breaks gauge symmetry in the A phase, where it also exhibits fourfold rotational symmetry in the *a-b* plane distinct from the hexagonal symmetry of the crystal lattice. In the B phase, the order parameter becomes isotropic in the *a-b* plane as $T \to 0$ K, and the phase difference between the real and imaginary components imparts an overall angular momentum to the pair wavefunction. Hence, time-reversal symmetry (TRS) is broken in this phase, with the sign of the imaginary component determining the orientation (chirality) of the internal angular momentum of the pair along $\pm\hat{z}$.

The results of Josephson interferometry experiments (*13, 14*) are consistent with the spatial symmetries of the $E_{2u}$ order parameter of Eq. (1). However, attempts to observe TRS-breaking (TRSB) in UPt$_3$ via μSR measurements have yielded conflicting results (*15, 16*). Moreover, recent thermal conductivity data (*17*) have been interpreted to support a gap function belonging to an $E_{1u}$ representation that precludes TRSB in the B phase. Thus, the unresolved question of whether TRS is indeed broken in the B phase has become critical to determining the symmetry and hence the proper classification of the superconducting order parameter of UPt$_3$.

A general consequence of a TRSB order parameter (with a net moment oriented along the *c* axis) in the presence of particle-hole asymmetry is the appearance of a finite difference between the complex indices of refraction for right ($n_R$) and left ($n_L$) circularly polarized light, resulting in ellipticity and circular birefringence (*12*). In particular, the polar Kerr angle, which measures the degree of rotation of linearly polarized light at normal incidence, is related to the off-diagonal terms of the optical conductivity by (*18*)

$$\theta_K = -\Im\left(\frac{n_R - n_L}{n_R n_L - 1}\right) \approx \frac{4\pi}{\omega d} \cdot \frac{\sigma''_{xy}(\omega)}{\tilde{n}(\tilde{n}^2 - 1)}, \tag{2}$$

where $\tilde{n}$ is the average index of refraction of the material, $d$ is a microscopic length scale, and $\omega$ is the frequency of the incident light. The off-diagonal term of the conductivity tensor, $\sigma_{xy} = \sigma'_{xy} + i\sigma''_{xy}$, is nonzero if TRS is broken (*19*), and the second equality holds for weak absorption (*21*). In the simplest estimate, $\sigma_{xy}$ will have the natural scale of the Hall conductivity $e^2/h$, reduced by a factor that reflects the large difference between the superconducting gap and probe energies: $\Delta_0 \ll \hbar\omega$ in the optical regime, where $\Delta_0$ is the magnitude of the order parameter at the Fermi level. Since the current response to the incident electric field involves a product of the wavefunction and its gradient, the conductivity must be proportional to $\Delta_0^2$, yielding $\sigma''_{xy}(\omega) \sim \frac{e^2}{hd}(\Delta_0/\hbar\omega)^2$ (*21*). As a result, the degree of optical rotation expected from TRSB in unconventional superconductors is expected to be on the order of 1 μrad or less (*12*). Measurements of such small rotations pose challenges to experiment; nevertheless, high-resolution measurements of PKE have been used to place limits on anyon superconductivity in the high-$T_c$ cuprates (*22*) and to establish TRSB in the order parameter of the spin-triplet superconductor $Sr_2RuO_4$ (*23*).

The requirement to resolve sub-μrad optical rotation without application of an external TRSB field (such as a magnetic field) prevents us from employing the modulation techniques commonly used to detect small signals in magnetic materials. Thus, to measure Kerr rotation in $UPt_3$, we have constructed a fiber-based zero-area loop Sagnac interferometer operating at 1550 nm (*24, 25*) with a focused spot size of 10.6 μm. The interferometer is designed and biased to reject optical signals due to nonreciprocal effects (e.g. linear birefringence and backscattering off of interfaces), allowing us to resolve changes in Kerr angle on the order of 50 nano-radians with a minimum of optical power incident on the sample. This latter quality is of particular significance in this study, as spot heating in the temperature range of interest (300-600 mK) can be substantial even for thermally well-anchored samples with high thermal conductivities. To optimize the signal to noise ratio (SNR), we typically applied 20 μW of optical power to the sample, of which roughly forty percent is expected to have been absorbed (*26*). Lower incident powers (with inferior SNR) were also used to verify our results (*27*).

Our experiments were performed on a 4.5×3.3×0.91 mm single crystal of $UPt_3$ with residual resistivity ratio RRR ~ 850 measured along the *c* axis (Fig. 1). The crystal was mounted with GE varnish to a copper stage attached to the cold finger of a helium-3 cryostat, with the optical probe beam incident on the *a-b* plane of the crystal. The sample was then cooled down to the cryostat's base temperature, and Kerr angle was measured in ambient magnetic field ($H_{ext} < 0.3$ Oe) as the sample was slowly warmed through $T_{c-}$ and $T_{c+}$. In addition to measuring Kerr angle, we installed a split-coil mutual inductance assembly (*28*) operating at 100 kHz with a drive amplitude of ~8.5 mG. Mutual inductance measurements of the sample could then be conducted under the same experimental conditions as the Kerr measurements in order to verify the upper transition temperature of ~550 mK (see Fig. 2).

The main experimental result is summarized in Figure 2. On the left-hand axis we plot the Kerr angle versus temperature obtained in a typical zero-field warmup following a zero-field cooldown. A temperature-independent background, consisting of electrical and optical offsets in the instrument, has been averaged between 600 and 700 mK and subtracted from the plotted data as in (*23*); the equivalent magnitude of this offset varied between ~0.2 and ~3 μrad over all of the data taken on UPt$_3$, although for most experimental runs the equivalent offset was of order 0.3 μrad (*27*). Above $T_{Kerr} \approx 460$ mK, the measured Kerr signal remains unchanged from the background. Below this temperature, however, a finite change in Kerr angle appears, which reaches ~0.4 μrad at 350 mK and falls to zero at $T_{Kerr}$. Finite element analysis using the thermal conductivity and heat capacity data of (*29, 30*) allows us to estimate the degree of sample heating to be no more than 50 mK at the spot of illumination. The resulting uncertainty in the transition temperature in the limit of zero optical heating is indicated by the thick gray bar in Fig. 2.

Theoretical treatments of the optical response of chiral superconductors suggest that the Kerr effect varies as the product of the real and imaginary components of the order parameter (*31*). The Kerr effect may arise either via impurity scattering (*32, 33*) or interband coupling (*34*), although for our measurements the latter mechanism is more likely given the high purity of our sample and the involvement of five bands in the Fermi surface (*1*). Because the real and imaginary components of the proposed order parameter for UPt$_3$ have different transition temperatures, the phenomenological expression for $\theta_K$ involves both $T_{c+}$ and $T_{c-}$:

$$\theta_K \propto \sqrt{\left[1-\left(T/T_{c+}\right)^2\right]\cdot\left[1-\left(T/T_{c-}\right)^2\right]}. \qquad (3)$$

This functional form is plotted as a solid line in Fig. 2 and is consistent with measurement, although the scatter in the data is too large to exclude the possibility of a more complex relationship between $\theta_K$, temperature, and the gap structure of UPt$_3$.

To compare the observed optical Kerr transition with the onset of A-phase superconductivity in UPt$_3$, a baseline magnetic susceptibility curve, taken with no incident light from the Kerr probe, is shown on the right-hand axis of the plot. We identify the upper superconducting transition temperature $T_{c+} \sim 550$ mK with the departure of the susceptibility from the normal state linear background, as in (*35*). The absorption of light from the Kerr measurement results in bulk sample heating that depresses the measured mutual inductance $T_c$ from the 0 µW curve by no more than ~8.5 mK for the optical powers used in this study (Fig. S3). This offset is indicated by the thin gray band in Fig. 2. The transitions at $T_{c+}$ and $T_{Kerr}$ are clearly separated in temperature; in the absence of any other structural or electronic phase transitions in this temperature range, we ascribe the additional broken symmetry at $T_{Kerr}$ to the onset of B phase superconductivity at $T_{c-}$.

A defining characteristic of spontaneous symmetry-breaking is the sensitivity of the order parameter to alignment by a small symmetry-breaking field as the transition temperature is crossed. In the case of UPt$_3$, the TRSB state is expected to couple to magnetic fields applied along the *c* axis (*12*); therefore it should be possible to train the sign of the TRSB Kerr signal with an arbitrarily small external field oriented in this direction. We show the results of field-training measurements of the Kerr effect in Figure 3. For these experiments the sample was cooled to base temperature in an applied field of +50 or -50 Gauss. At base temperature the field was switched off, and Kerr angle was again measured upon warmup. The residual magnetic field with the magnet off was identical to that of the zero-field cooled measurements to within the resolution of a commercial Hall sensor installed in the sample space.

We note several important features in the data. First, the signal tracks the direction of the training field, showing that the TRSB order parameter couples to magnetic fields as expected. Second, the maximum size of the field-trained signal matches that of the zero-field measurements to within our experimental resolution. While in general, one would expect TRSB domains to form with random alignments, leading to partial or even complete cancellation of Kerr signal in the area sampled by the probe beam, our repeated observation of maximal Kerr signal even in the absence of field training is consistent with observations elsewhere that UPt$_3$ tends to spontaneously form very large superconducting domains, possibly spanning the entire crystal (*13*, *14*, *17*). Finally, the absence of a discernable additional Kerr signal in the field-cooled measurements implies that the TRSB signal originates from the superconducting order parameter itself, rather than from vortices induced by an external field.

To properly analyze broken TRS in the superconducting state, one must also consider the potential influence of magnetic effects already present at higher temperatures. In UPt$_3$, antiferromagnetic (AF) correlations associated with magnetic moments on the uranium sites appear at 20 K, with static in-plane AF order developing below $T_N$ ~ 5 K (*7*, *36*). TRS is therefore broken well above $T_{c+}$ and $T_{c-}$. However, one can still examine the superconducting condensate for additional TRSB that is ferromagnetic in character, that is, displaying a two-state degeneracy due to the presence of an imaginary component in the gap function as discussed above. This additional out-of-plane signal can be considered independently from the in-plane TRSB arising from antiferromagnetism, although the AF order may still couple to superconductivity by acting as, e.g., a symmetry-breaking field (*7, 12*). We can verify the independence of these two sources of TRSB by measuring Kerr rotation through $T_N$. As expected for in-plane ordering with no net moment, we find no signature (within ±0.1 μrad) of the AF transition in the Kerr effect under both zero-field and field-cooled conditions (Fig. S4), despite the enhancement provided by the large spin-orbit coupling present in the material (*12*). The null result at $T_N$ suggests that the signal observed in the B phase is a true property of superconductivity in UPt$_3$, rather than a secondary effect of the background AF ordering in this system.

The appearance of broken TRS in the B phase carries with it several implications for the theory of superconductivity in UPt$_3$. Because a general multicomponent superconducting order parameter can break TRS only if it belongs to a multidimensional representation (*37*), our results taken alone imply that the superconducting order parameter belongs to one of the four two-dimensional representations of $D_{6h}$. In principle, the data are consistent with any choice of basis functions with relatively complex coefficients within these representations. However, the preponderance of other experimental evidence (*1, 9-11, 13-15, 30*) narrows the possibilities further, leaving the $E_{2u}$ order parameter of Eq. 1 as the likely candidate to describe the superconducting phases of UPt$_3$. Although a microscopic theory for polar Kerr effect in UPt$_3$ is not available at present, the temperature dependence of our data appears to be consistent with the phenomenology developed in Ref. (*34*) to account for a finite Kerr rotation in a multiband superconductor with interorbital coupling $\varepsilon_{12}$ and intraband pairing: $\theta_K \propto \varepsilon_{12}\eta_1\eta_2$. This in turn suggests that Kerr effect measurements can more generally be used to track the temperature dependence of the gap function in TRSB superconductors.

*************************

**Acknowledgments:** Stimulating discussions with S. Kivelson, A. Huxley, and D. Agterberg; sample characterization by K. Avers (Northwestern); and instrument design assistance from G. Burkhard (Stanford) are greatly appreciated. This work was supported by the U.S. Department of Energy Office of Basic Energy Science, Division of Materials Science and Engineering, at Stanford under contract no. DE-AC02-76SF00515 and at Northwestern under contract no. DE- FG02-05ER46248. Construction of the Sagnac apparatus was partially funded by the Stanford Center for Probing the Nanoscale (NSF NSEC 0425897). ERS received additional support from a Gabilan Stanford Graduate Fellowship and the DARE Doctoral Fellowship Program.

*************************

**Supplementary Materials:**

Materials and Methods

Further discussion of mutual inductance, Kerr effect, and sample heating

Figures S1-S4

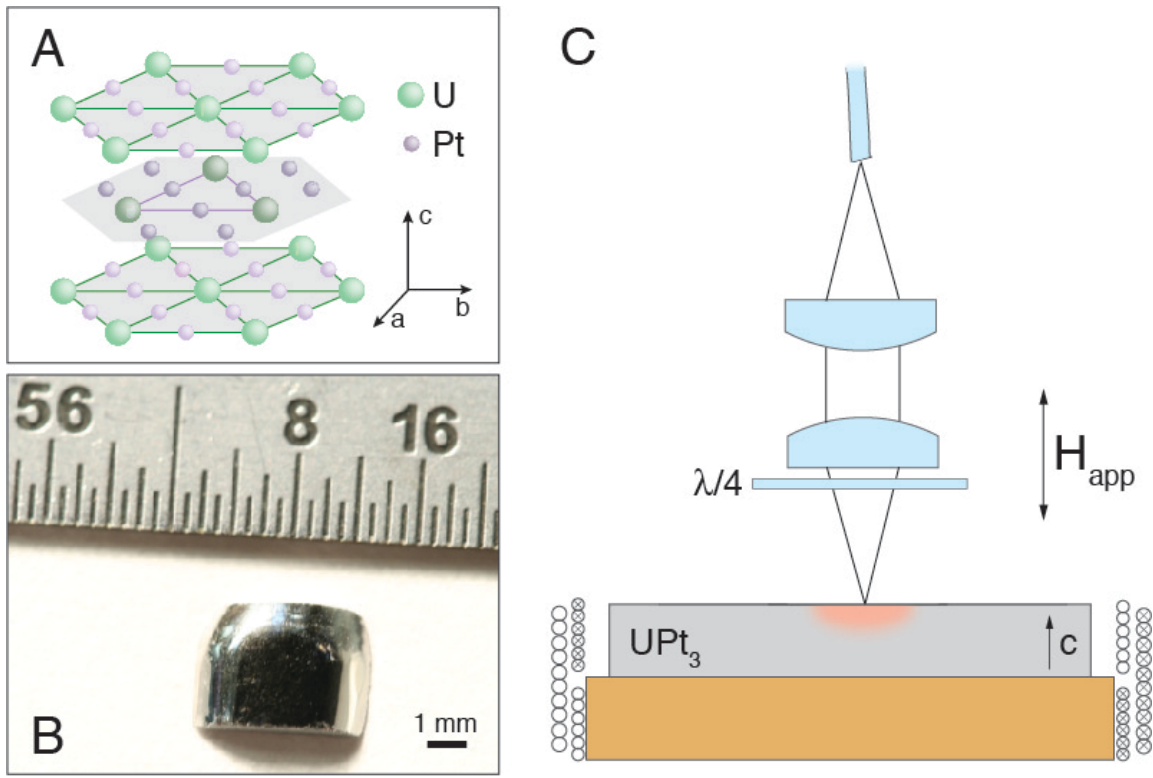

**Fig. 1. Material information and experimental setup.** (**A**) Crystal structure of UPt$_3$ and (**B**) photograph of the single crystal used in this study. Length scale on ruler is in sixty-fourths of an inch. (**C**) Experiment geometry of the Kerr rotation and mutual inductance measurements.

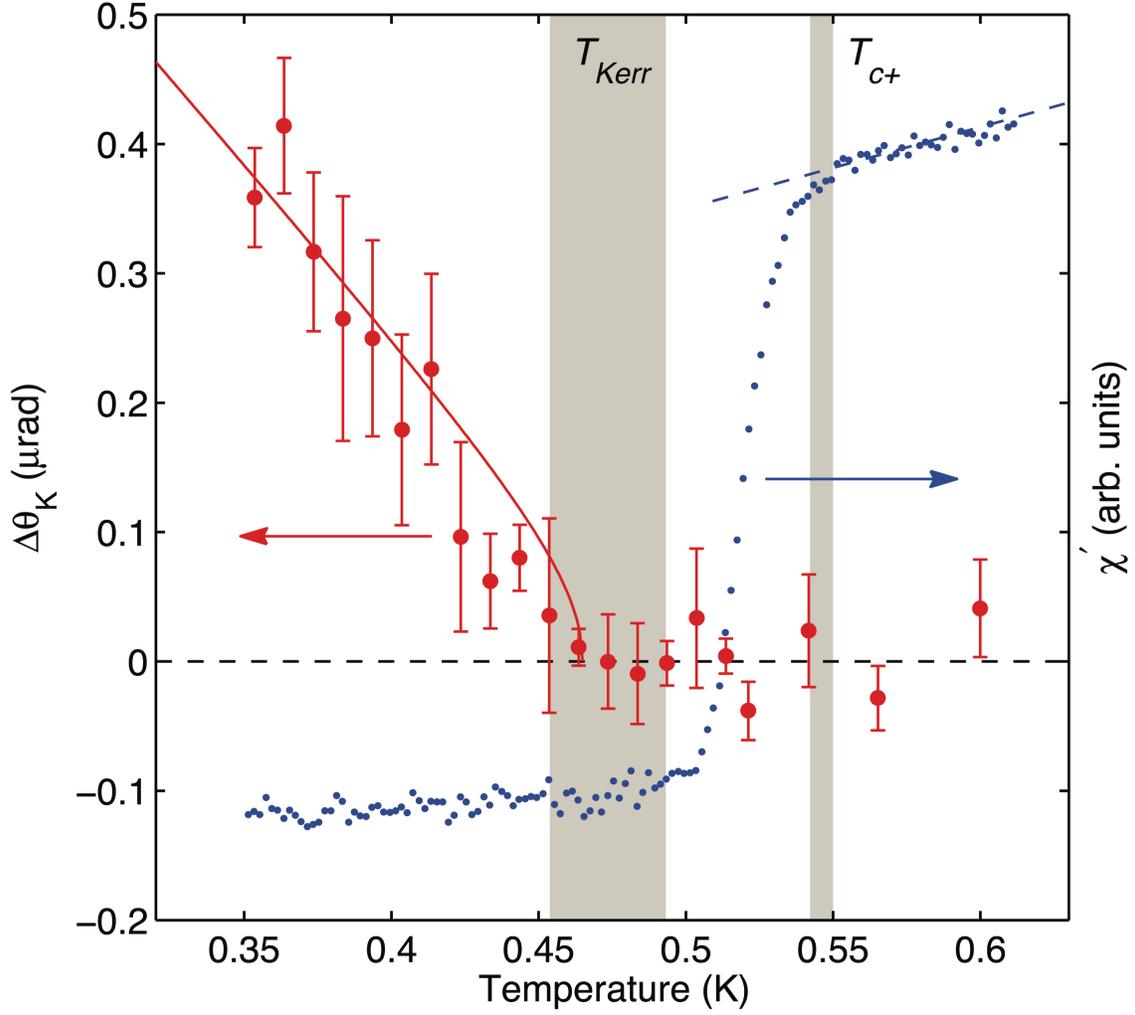

**Fig. 2. Measurement of Kerr effect and $T_{c+}$ in UPt$_3$.** Kerr angle (red, left axis) and the real part of the mutual inductance (blue, right axis) are plotted as a function of temperature for a UPt$_3$ single crystal. The onset of superconductivity at $T_{c+}$ is well removed from the onset of TRSB at $T_{Kerr} \sim 460$ mK $\sim T_{c-}$. The gray bands indicate uncertainties in $T_{c+}$ and $T_{c-}$ from the effects of optical heating. Error bars on the Kerr data are statistical. Solid line is a guide to the eye of the form

$$\theta_K \propto \sqrt{\left[1-(T/T_{c+})^2\right]\cdot\left[1-(T/T_{c-})^2\right]}.$$

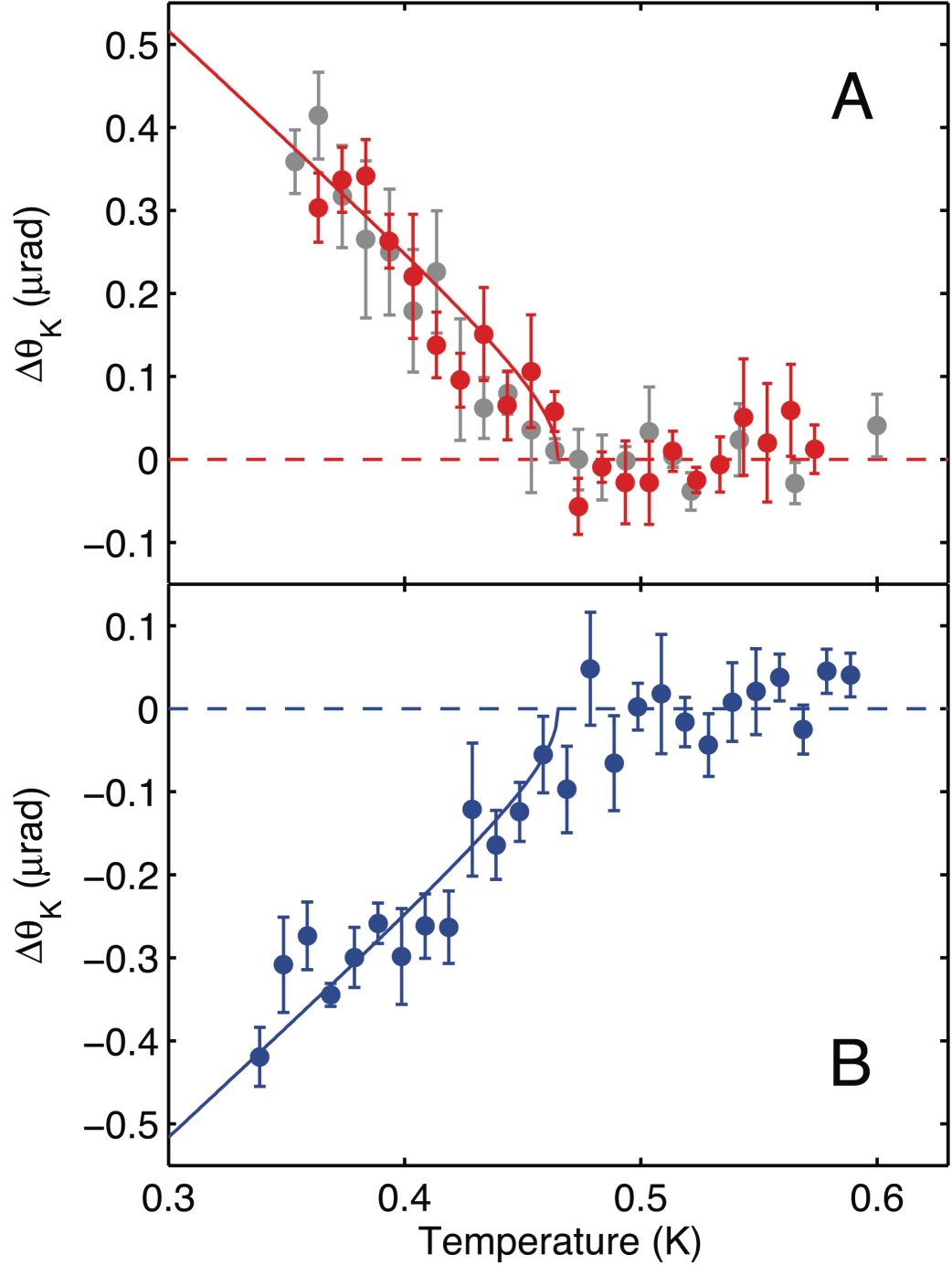

**Fig. 3. Magnetic field training of the Kerr effect.** Solid lines are guides to the eye of the form $\theta_K \propto \sqrt{\left[1-(T/T_{c+})^2\right]\cdot\left[1-(T/T_{c-})^2\right]}$. **(A)** Zero-field warmup data after cooling the sample through $T_{c-}$ in a +50 G field (red) and in zero field (gray, replotted from Fig. 2). **(B)** Zero-field warmup data after cooling the sample down in a -50 G field, showing complete reversal of the TRSB signal.

## Supplementary Materials:

Materials and Methods

Further discussion of mutual inductance, Kerr effect, and sample heating

Figures S1-S4

*************************

## Materials and Methods

### Sample growth and preparation

The UPt$_3$ crystal used in this study was grown by electron-beam float-zone refining. Several facets on the surface of the as-grown mother crystal were identified to have the *c*-axis normal. One such facet, approximately 2.5×2.5 mm$^2$, was cut from the mother crystal for this experiment, with the as-grown surface preserved. The crystal was annealed for one week at 850 degrees C with slow warming and cooling. A whisker cut from the mother crystal directly adjacent to this sample was measured to have a residual resistivity ratio (RRR) of ~850, upper transition temperature $T_{c+}$ = 553 mK (±2 mK), and resistive transition width $\Delta T_c$ = 9 mK (±1 mK).

### Measurement apparatus

High-resolution measurements of Kerr rotation were performed using a fiber-based zero-area loop Sagnac interferometer [for a thorough review, see Ref. (*23*)]; a schematic of the apparatus used in this study is shown in Fig. S1 below. The two linearly polarized counterpropagating beams comprising the interferometer are isolated along the fast and slow axes of ~10 m of polarization-maintaining fiber and are modulated in phase by $\phi_m \sin(\omega t)$, and $\phi_m \sin[\omega(t+\tau)]$, respectively, where $\tau$ is the transit time of light through the interferometer, $\omega = \pi/\tau \approx 5.175$ MHz is the modulation frequency, and the modulation amplitude $\phi_m \sim 0.92$ rad is chosen to maximize the resolution of the interferometer. The linearly polarized light traveling along each of these paths passes through a quarter-waveplate and is converted into circularly polarized light just above the sample. Upon reflection, then, one branch of the interferometer acquires a phase shift of $+\theta_K$, while its orthogonally polarized counterpart acquires an opposite phase shift of $-\theta_K$. Under these conditions the detected intensity $I(t) = |E_{tot}(t)|^2$ can be decomposed into harmonics of the modulation frequency:

$$\frac{I(t)}{I_0} = \frac{1}{2}[1+J_0(2\phi_m)] - $$
$$-[\sin(2\theta_K)J_1(2\phi_m)]\sin(\omega t) + [\cos(2\theta_K)J_2(2\phi_m)]\cos(2\omega t) + ... , \quad (S1)$$

where $J_0$, $J_1$, and $J_2$ are Bessel functions of the first kind. One can then extract the Kerr angle via lockin detection of the first and second harmonic amplitudes at the photodetector, using the relation

$$\theta_K = \frac{1}{2}\tan^{-1}\left[\frac{J_2(2\phi_m)}{J_1(2\phi_m)} \cdot \frac{V_\omega}{V_{2\omega}}\right]. \tag{S2}$$

The probe end of the fiber was installed through a vacuum fitting into a Janis model HE-3-SSV $^3$He cryostat for low temperature measurements.

Data analysis

The raw data collected for eventual determination of the Kerr angle consists of the components of the photodetector output voltage that vary as the first harmonic ($V_\omega$) and second harmonic ($V_{2\omega}$) of the modulation frequency $\omega$ (Eq. S2). For the sample reflectivity and incident optical powers used this study, the measured $V_\omega$ was typically on the order of ≤1 µV, while $V_{2\omega}$ was several orders of magnitude greater: $V_{2\omega} \sim$ 10-20 mV. An accurate determination of the Kerr angle therefore relies sensitively on the analysis of the first harmonics channel, where instrumentation offsets on the sub-µV level may generate a significant background.

We approach the problem of identifying and subtracting this background by assuming that the measured (small) first harmonics voltage contains contributions both from the sample and from instrument offset: $V_\omega = V_{\text{Kerr}} + V_{\text{offset}}$. The $V_{\text{offset}}$ term comprises offsets due to electrical background and optical misalignments. Because some optical components and their mounts reside inside of the $^3$He cryostat (Fig. S1), the total offset may assume a slight temperature dependence; however, in the temperature ranges of interest this dependence may be ignored as we show below. In Figure S2 we plot the ratio $V_\omega/V_{2\omega}$ for a data set with a relatively large background. The sharp change in first harmonics voltage due to onset of Kerr effect at a known thermodynamic phase transition is easily distinguishable from the background, which will vary smoothly (if at all) with temperature and/or time. It is important to note that, once normalized by $V_{2\omega}$, the signal generated by the Kerr effect has the same magnitude regardless of the size of the background on which it sits.

The final data are binned by temperature range and averaged to generate the figures in the main text. Error bars represent the standard deviation of data points within each temperature bin.

**Supplementary Text**

<u>Further discussion of mutual inductance, Kerr effect, and sample heating</u>

  A thorough characterization of sample heating due to absorption of light from the Kerr probe is critical for determining the relationship between the optical Kerr-transition temperature $T_{\text{Kerr}}$ and the superconducting transition temperatures $T_{c+}$ and $T_{c-}$. Fig. S3 shows the effect of incident optical power on the bulk transition temperature, as measured by mutual inductance (MI). For increasing optical powers, the mutual inductance transition is shifted to lower temperatures. To quantify this effect, we determine $T_{c+}$ for each MI curve by taking the difference in temperature for the midpoint of each transition relative to the midpoint of the 0 μW transition and reference this to the onset temperature $T_{c+}(P_{\text{incident}} = 0\ \mu W) \sim 550$ mK. The result, plotted in the inset, shows the linear dependence of $T_{c+}$ on incident optical power, with a maximum depression of ~8.5 mK for the highest optical power (20 μW) used in this study to measure Kerr effect.

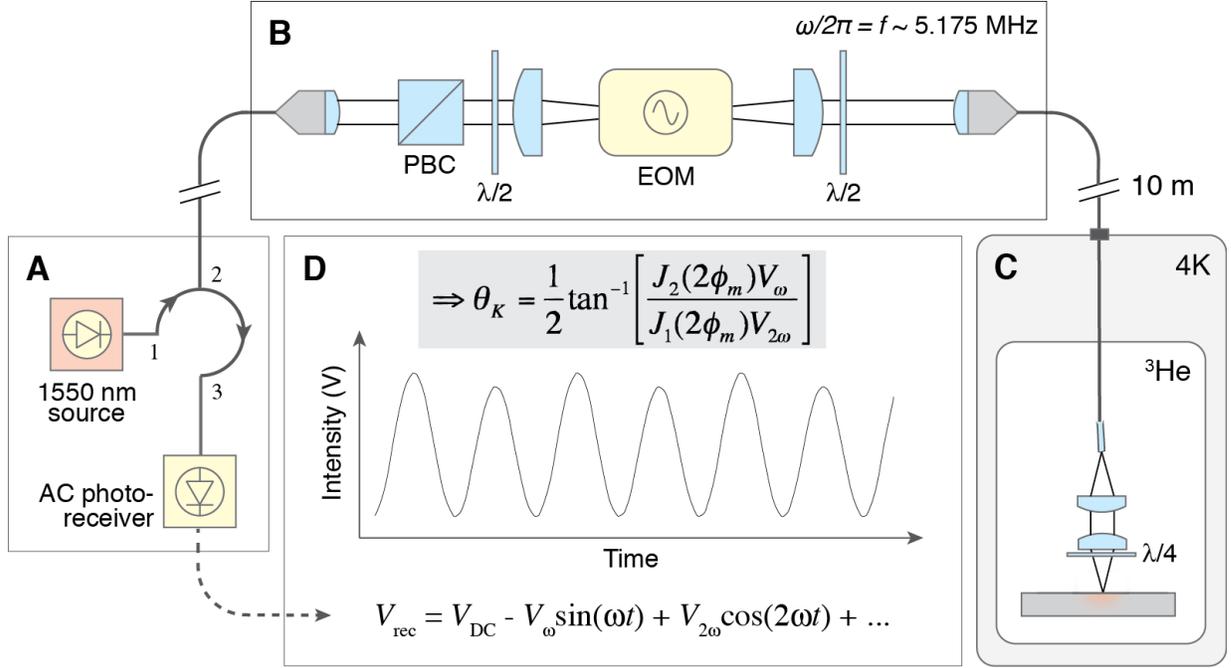

**Fig. S1. Schematic of the zero-area loop Sagnac interferometer used in this study.**
**(A)** Light from a broadband (superluminescent LED) source centered at λ=1550 nm is launched into port 1 of a polarization-maintaining circulator, where it enters the interferometer via port 2. Returning light enters at port 2 and is detected at port 3. **(B)** Bulk optics of the interferometer. Light from the 1550 nm source is linearly polarized at 45°; the vertical component travels along the extraordinary axis of the electro-optic modulator (EOM) and is phase-modulated by $\phi_m\sin(\omega t)$, while the horizontal component travels along the ordinary axis and does not undergo phase modulation. One component is launched into the fast axis of ~10 m of polarization-maintaining fiber, while the other is effectively isolated on the slow axis. **(C)** Linearly polarized light traveling along the fast and slow axes of the fiber is routed into a He-3 cryostat, where it is focused and converted to circularly polarized light before reaching the sample. Upon reflection, one circular component experiences a phase shift of $+\theta_K$, while the other experiences a phase shift of $-\theta_K$. After a second pass through the quarter waveplate, each component is launched back into the fiber along the slow and fast axes, respectively. The light that initially traveled along the ordinary axis of the EOM now returns along the extraordinary axis and is phase-modulated by $\phi_m\sin(\omega(t+\tau))$, where $\tau$ is the transit time of light through the interferometer. **(D)** The resulting interference pattern can be decomposed into harmonics of the modulation frequency; the Kerr angle is then proportional to the ratio of the first and second harmonic amplitudes.

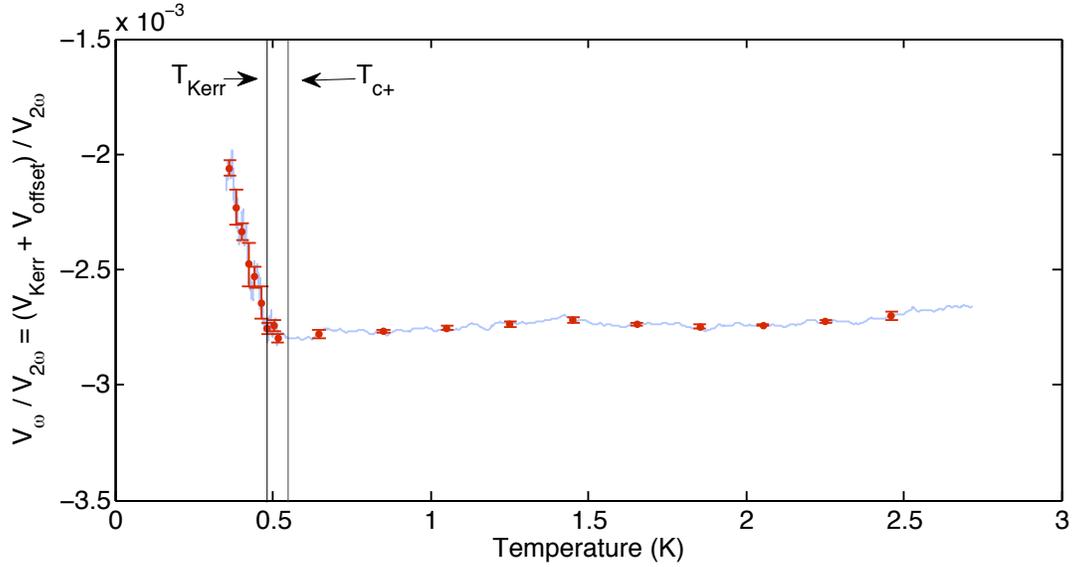

**Fig. S2. First harmonics voltage data normalized by second harmonics, showing Kerr signal and instrument background.** Light blue is the as-measured $V_\omega/V_{2\omega}$ with a twenty minute time constant; red are the same data averaged over temperature bins of width 0.05 K below 0.5 K and 0.2 K above 0.5 K. The sharp onset of signal due to Kerr effect in the B phase is distinct from the smooth background generated by first harmonic voltage offsets above $T_{Kerr}$. Having established that there is indeed a feature in the data below ~0.5 K, higher resolution (i.e. slower temperature ramp rate) data is typically taken below ~0.6 K to determine $T_{Kerr}$ more exactly.

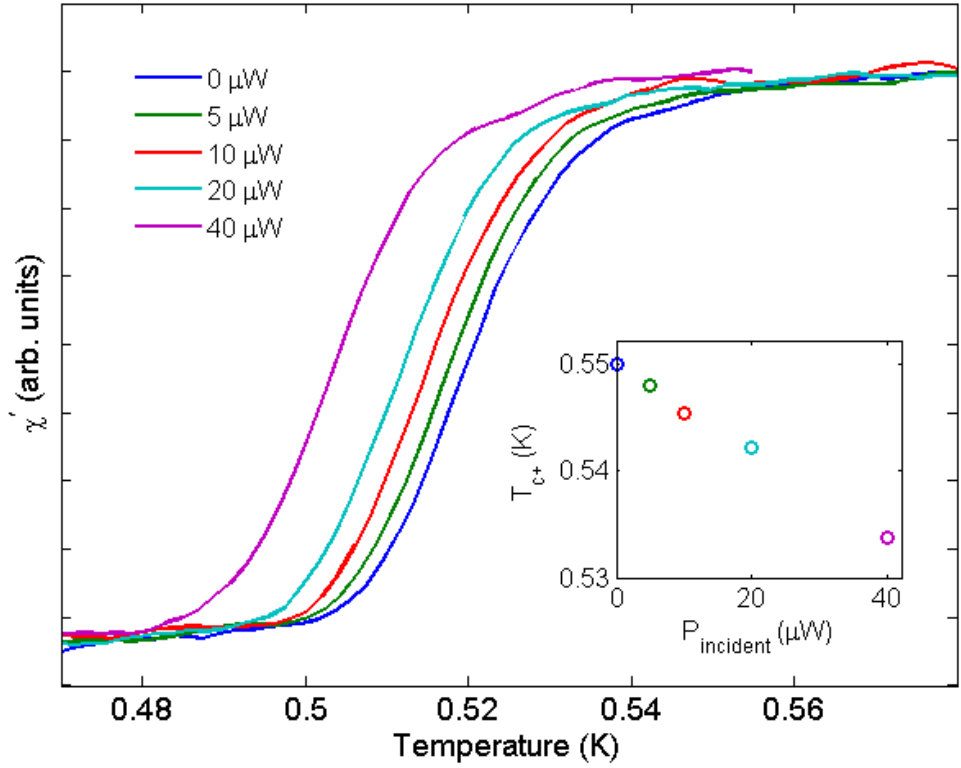

**Fig. S3. Effect of sample heating due to incident light on mutual inductance (MI) data.** Solid lines: real (in-phase) component of MI signal at various incident optical powers. Optical heating of the crystal is evidenced by the downward shift in the MI data with increasing incident power. Inset: bulk $T_{c+}$ estimated by referencing the shift in midpoint of the MI transition to the onset of the transition for 0 μW incident power.

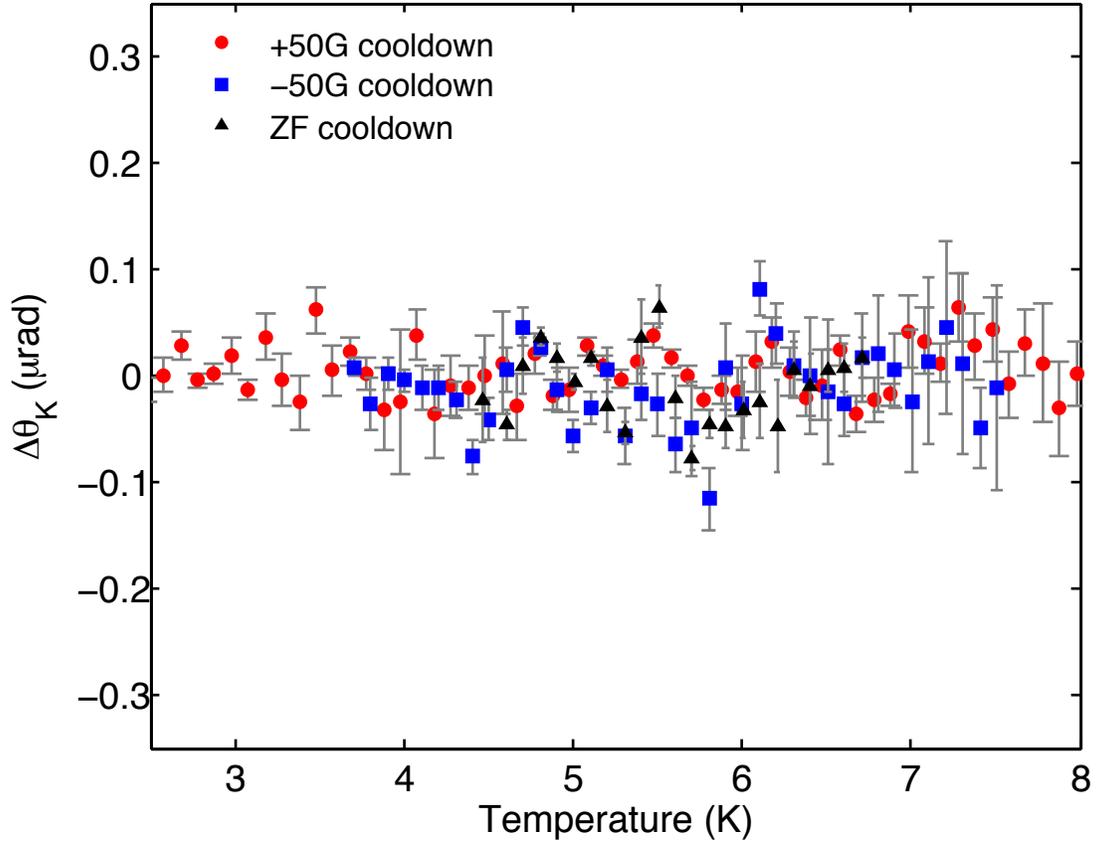

**Fig. S4. Kerr effect data taken on warmup through the Néel transition $T_N \sim 5$ K.** The sample was cooled to base temperature in +50G (red circles), -50G (blue squares), and zero (black triangles) applied field. Error bars are suppressed in gray for clarity. We do not resolve any influence of the in-plane antiferromagnetic transition on the measured (out of plane) Kerr signal to within ±0.1 μrad, suggesting that the antiferromagnetic transition has no direct effect on the TRSB signal observed in the superconducting B phase.